# Spin-state Directed Synthesis of >20 μm 2D Layered Transition Metal Hydroxides via Edge-on Condensation

Lu Ping[1]*, Gillian E. Minarik[2]*, Hongze Gao[2], Jun Cao[2], Tianshu Li[1], Hikari Kitadai[2], Xi Ling[1,2,3]†

[1] Division of Materials Science and Engineering, Boston University, 15 St. Mary's Street, Boston, MA 02215, USA.

[2] Department of Chemistry, Boston University, 590 Commonwealth Avenue, Boston, MA 02215, USA.

[3] The Photonics Center, Boston University, 8 St. Mary's Street, Boston, MA 02215, USA.

*These authors contributed equally to this work.

†Corresponding author. Email: xiling@bu.edu

**Abstract**

Layered transition metal hydroxides (LTMHs) with transition metal centers sandwiched between layers of coordinating hydroxide anions have attracted considerable interest for their potential in developing clean energy sources and storage technologies. However, two dimensional (2D) LTMHs remain largely unstudied in terms of their physical properties and the applications in electronic devices. Here, directed by the relationship of the spin state of 3d transition metal (TM) ions such as Ni, Co, Cu, and the corresponding geometry of the crystal field, we discover that $Ni^{2+}$ with perfect $O_h$ symmetry is ideal for intraplanar growth, leading to the achievement of >20 μm α-$Ni(OH)_2$ 2D crystals with high yield, which are the largest 2D domains reported so far. We also report the successful synthesis of 2D $Co(OH)_2$ crystals (>40 μm) with less yield due to the slight geometry distortion resulted from uneven number of electrons. Moreover, the detailed structural characterization of synthesized α-$Ni(OH)_2$ are performed; the optical band gap energy is extrapolated as 2.54 eV from optical absorption measurements and is measured as 2.50 eV from reflected electrons energy loss spectroscopy (REELS), suggesting the potential as insulating 2D dielectric material for electronic devices. Furthermore, key parameters of the hydrothermal reaction including soaking temperature, starting pH and cooling rate, are systematically tuned to understand their effects on morphological and crystallographic perspectives, allowing the establishment of a 2D growth mechanism. This work demonstrates a scalable pathway to synthesize large 2D LTMHs from simple methods, paving the way for the study of fundamental physical properties and device applications of 2D LTMHs.

# Introduction

The first successful exfoliation of graphene in 2004 sparked a dramatic increase in 2D materials research and the repertoire of reported 2D materials family has since expanded. 2D crystals composed of single- or few-layers of atoms, display extraordinary chemical, optical, and electronic properties compared with their bulk 3D counterparts due to the electron and phonon confinement effect in the 2D limit[1]. Nevertheless, only several 2D materials such as graphene[2], transition metal dichalcogenides[3,4] and hexagonal boron nitride[5], are reasonably well-studies for electronic and optoelectronic applications, thanks to the development of successful synthesis methods for large domains and films. Among the over thousands of van der Waals 2D materials,[6,7] unfortunately, many of them are underexplored experimentally, due to the lack of proper synthesis methods for sufficiently large domains. Layered transition metal hydroxides (LTMHs) are a group of materials with van der Waals layered structures, approaches through chemical reactions such as chemical precipitation[8–11], electrochemical precipitation[12,13], chemical aging[14–16], sol-gel[17], hydrothermal[18–21], solvothermal[22] methods, have been widely attempted to synthesize 2D LTMHs. Low surface energy allows the physical exfoliation to be applied to exfoliate 2D sheets from bulk under mild conditions.[17,23] The obtained materials have demonstrated capability as high-performance electrocatalysts for oxygen evolution reaction (OER)[24,25] and high-capacity electrode materials for supercapacitors[26,27], thereby exhibiting great potential for applications in terms of clean energy[28–30] and energy storage[31–34]. However, the reported 2D LTMHs are usually nanosheets[17,23], nanoplates[20,35], nanoflowers[36] with domain size from hundreds of nm to a few μm, larger 2D LTMHs (>10 μm of lateral size) that are desired for certain applications such as semiconductor devices have never been reported[37,38]. Thus, studies of their electronic and magnetic properties have been limited to theoretical investigation[39–41].

In this work, we achieve largest ever single crystals of 2D α-Ni(OH)$_2$ (>20 μm) and Co(OH)$_2$ (>40 μm), guided by the investigation of spin-state directed geometry of first row 3d transition metal ions(Figure 1). [Ni(H$_2$O)$_6$]$^{2+}$ which has the perfect O$_h$ symmetry, is stable and preferable to add on more crystal units along intraplane direction, allowing the realization of large 2D domains with large yield. For Co$^{2+}$, Cu$^{2+}$, Fe$^{2+}$, due to the imbalanced distribution of electrons on the degenerated orbitals, their six-coordination suffer from different extent of distortion, thus 2D intraplane grown is not anticipated (Cu$^{2+}$, Fe$^{2+}$) or with less yield (Co$^{2+}$). Besides, a 2D growth mechanism of LTMHs at atomic level is established and a comprehensively experimental study is performed using hydrothermal method. Further characterizations on the structure and dielectric properties of the synthesized 2D Ni(OH)$_2$ reveal the high crystallinity, primary α phase, and a bandgap of 2.54 eV with a high dielectric strength of 3.2 MV/cm on a 4.9 nm flake, showing a great potential as gate dielectric materials for 2D field effect transistors (FETs) devices. Due to the scalability of hydrothermal method, we anticipate this achievement will pave the way for the fundamental property study of 2D LTMHs crystals that is not accessible on samples with small domains and trigger the interests in exploring their applications in 2D micro-nanoelectronics devices.

**Results and Discussion**

**Relation of spin-state and crystal field geometry of 3d transition metal ions**

To form 2D crystals, isotropic growth along the intraplanar direction is required. We start with looking into the geometries of solvated transition metal ions, which macroscopically determines the resultant crystal structure and morphology. The spin-state of transition metal complexes dictates their geometries as outlined by the Jahn-Teller (JT) theorem, which describes spontaneous symmetry breaking to provide energetic stabilization of transition metal complexes when there is an odd number of electrons occupying degenerate orbitals, thus determines the geometries of transition metal complexes.[42] The JT effect persists in both solution and the solid state, and has long been used to understand the geometrical structures of molecules and crystals.[43,44] Then we examine the simple case of high symmetry six-coordinate, homoleptic monomer $[M(H_2O)_6]^{2+}$ [M=Ni, Co, Cu] (Fig. 1), which forms from the dissolution of precursors and matches the (pseudo)octahedral field seen by the metal ions in the LTMH lattice. Since $H_2O$ and $OH^-$ are weak field ligands, the complexes are high-spin.

Specifically, $Ni^{2+}$ with its $d^8$ configuration (Fig. 1A) has even number of electrons on degenerated orbitals, experiencing no JT distortions and giving rise to perfect $O_h$ symmetry on $[Ni(H_2O)_6]^{2+}$ with the same bond length and same electron density on all six $Ni-H_2O$ metal-ligand coordination bonds. This geometry is ideal and stable for more $[Ni(H_2O)_6]^{2+}$ units to add on via the edge-on incorporation from any direction, thus the crystal growth is isotropic along intraplanar direction, leading to circular 2D domains. For $[Co(H_2O)_6]^{2+}$ with high spin $d^7$ configuration (Fig 1B), it suffers slight JT distortion and has weak compression along z-axis due to one vacancy in $t_{2g}$ set, leading to $D_{4h}$ symmetry.[45] The small degree of JT distortions reduces the number of ways the $[Co(H_2O)_6]^{2+}$ units can add onto the exiting seed crystal with good structural alignment, resulting in distorted-quasi circular morphology. A more extreme case of JT distortion is presented with $[Cu(H_2O)_6]^{2+}$ (Fig. 1C), where the high spin $d^9$ configuration of $Cu^{2+}$ leads to three electrons in the $e_g$ set, and thus largely elongated along the z-axis.[45,46] While the elongated axial bonds in $[Cu(H_2O)_6]^{2+}$ also give the complex $D_{4h}$ symmetry, the degree of distortion is more pronounced owing to the fact that the $e_g$ set are anti-bonding orbitals, which in short, directs 3D growth yielding bulk crystals with prism-like morphology (Supplementary Fig. 1C). This also suggests that suppression of JT effects in metal complexes where they are more pronounced can be the key to extending the repertoire of 2D LTMHs beyond Ni(OH)$_2$ and Co(OH)$_2$.

**Structural Characterization of 2D α-Ni(OH)$_2$ and Co(OH)$_2$**

Fig. 2A-B show the crystal structure of α-M(OH)$_2$ (M=Ni and Co), which is a typical layered material with a "sandwich" structure of bivalent metals between two layers of octahedral-coordinated OH$^-$ anions. The intercalation, e.g., H$_2$O, expanded the interlayer space and distinguished α phase from another commonly existed polymorphs, β phase, which has a smaller interlayer gap due to the absence of intercalation (Supplementary Fig. 2).

Figure 2C shows the typical optical image of Ni(OH)$_2$ thin flakes synthesized under 120 °C for 18 hours with a starting pH=5.57 and a 1.5 °C/min cooling rate, where micrometer-size domains and circular shape are widely observed. The average domain size and thickness is 20 μm and 16 nm, respectively, through surveying one hundred randomly chosen flakes (Supplementary Fig. 3). Particularly, over 90% flakes are larger than 10 μm. Notably, this is the largest domain size of 2D Ni(OH)$_2$ ever reported, which is also extraordinary compared to other 2D nanosheets-morphology typically synthesized using hydrothermal methods.[19,47–50] Fig. 2D shows the atomic force microscope (AFM) image of a flake of 6.3 nm, corresponding to ~8 atomic layers based on the theoretical interlayer distance of α-Ni(OH)$_2$, 8.0 Å[51]. The surface morphology of the flakes is further characterized using scanning electron microscopy (SEM) (Fig. 2C inset). Although subtle dis-uniformity is observed, indicating slight roughness, there is no visible holes and cracks, suggesting the flakes are continuous and smooth in general. We also measured two perpendicular diameters for each flake on the intraplane orientation. The ratio of two perpendicular diameters taken for each flake is defined and calculated as aspect ratio, which is dimensionless. The plotted distribution (Supplementary Fig. 2A inset) indicates about 50% of flakes have an aspect ratio as

1.0, consistent with the circular shape observed visually and the perfect octahedral geometry of Ni(OH)$_2$, while 84% of surveyed flakes maintain quasi-circular dimensions with aspect ratios within the range 0.8-1.2, suggesting isotropic, outward radial crystal growth. Furthermore, the surface chemical state of synthesized Ni(OH)$_2$ flakes was analyzed by X-ray photoelectron spectroscopy (XPS). The survey spectra (Supplementary Fig. 4A) clearly indicated the occurrence of Ni and O elements at their binding energies.[52] The core-level spectrum of Ni 2p (Fig. 2E) showed the two major peaks at 856.1 (Ni 2p$_{3/2}$) and 873.8 eV (Ni 2p$_{1/2}$) with a spin energy separation of 17.6 eV, which is characteristic of the Ni$^{2+}$ in Ni(OH)$_2$.[53,54] Meanwhile, the O 1s spectrum (Supplementary Fig. 4B) was fitted into two peaks, located at 531.0 and 532.1 eV, representing Ni-O-H and H-O-H, respectively.[55] The high intensity of H-O-H also suggests the high composition of H$_2$O, mainly exist as intercalation between layers to form α-Ni(OH)$_2$.

Using similar synthesizing conditions, we also successfully obtained Co(OH)$_2$ thin flakes with even larger domains, but the shapes are random rather than circular and the number of large domains is significantly lower than that of Ni(OH)$_2$ (Fig. 2F). The observed flakes exhibit over 40 μm domain at some directions, which is also the largest Co(OH)$_2$ ultrathin domains ever reported.[56–58] The surface morphology revealed by SEM (Fig. 2F inset) suggests the continuity of these crystals and showed the edge of Co(OH)$_2$ thin flakes are irregular, which is consistent with its distorted geometry caused by the repulsion on $d_{x^2-y^2}$ direction. Furthermore, AFM measurements on a large Co(OH)$_2$ flake showing the thickness of 2.5 nm. XPS was applied to examine the surface chemical state of synthesized Co(OH)$_2$, while survey spectrum (Supplementary Fig. 4C) clearly indicated the occurrence of Co and O elements at their binding energies. The core-level spectrum of Co 2p (Fig. 2H) showed the two major peaks at 779.9 (Co 2p$_{3/2}$) and 795.6 eV (Co 2p$_{1/2}$) with a characteristic spin energy separation of 15.7 eV.[59]

To determine the crystallographic structure of the obtained Ni(OH)$_2$ thin flakes, atomic-resolved transmission electron microscopy (TEM) and powder X-ray diffraction spectroscopy (PXRD) are carried out and the experimental results are in great agreement with the theoretical crystal structure as we showed in Figure 2A-B. The low-resolution TEM image in Fig. 3A reveals the 2D nature of a micrometer large domain. The selected area electron diffraction (SAED) (Fig. 3A inset) reveals the hexagonal lattice structure as shown in the top view of crystal structure (Fig. 2B). The six sharp dots from SAED not only suggest the high crystallinity but correspond to diffraction of ($1\bar{2}0$), ($2\bar{1}0$), (110), ($\bar{1}20$), ($\bar{2}10$) and ($\bar{1}\bar{1}0$) planes, which match the simulated lattice pattern well (Supplementary Fig. 5A).[60] Moreover, high-resolution TEM image (Fig. 3B) resolves the atomic arrangement in the crystal, which allows the precise measurement of the spacing between (301) planes [i.e. d (301)=1.50 Å]. In addition, the selected-area height profile (Supplementary Fig. 5B) indicates the distance crossing over 12 lattice planes of (301) is 1.85 nm, corresponding to a d (301) of 1.54 Å. Notably, d (301) derived from TEM is in excellent agreement with that from the ideal crystal structure (i.e., 1.50 Å). Further structural and crystallographic insights are revealed by PXRD spectrum taken from the bulk material, formed from aggregates of the thin flakes. As shown in Fig. 3C, all characteristic peaks along the PXRD spectrum (red curve) are consistent with α-Ni(OH)$_2$,[51] the corresponding peak position (2θ), lattice planes and d-spacings are tabulated in Supplementary Tab. 1. According to Bragg's Law, $2d\sin\theta = n\lambda$, where n = 1, λ = 1.54184 Å, the peak at 2θ = 12.5 ° corresponds to a d-spacing of 7.08 Å, matching well with the theoretical value of 8.0 Å of the (001) plane spacing of the α phase rather than 4.6 Å of β phase.[51] For a trigonally symmetric unit cell for brucite and many isostructural TMLHs such as Ni(OH)$_2$, d (001) provides a metric for interlayer distance and strength of the interaction between 2D sheets. The ~ 1 Å contraction of calculated interlayer spacing from the theoretical value may be due to the partial interstratification of β-phase space (Schematic is shown in Supplementary Fig. 6), or less homogenous intercalation in general. Indeed, the unit cell dimensions of α-Ni(OH)$_2$ are extremely sensitive to method of preparation, owing to varying degrees of intercalated species, strain, and ionic impurities.[61] The peaks 2θ = 22.0 °, 34.2 °, 38.0 ° and 60.0 °, which are assigned to (002), (110), (111) and (301) planes of the α-Ni(OH)$_2$, respectively, and the corresponding d-spacing are 4.04 Å, 2.62 Å, 2.37 Å and 1.54 Å, respectively. Notably, the obtained d (301) (i.e., 1.54 Å) from XRD measurement is highly consistent with that from TEM measurement (i.e., 1.50 Å from high resolution TEM and 1.54 Å from area height profile) and the ideal crystal structure (i.e., 1.50 Å). Although the interlayer gap of the synthesized α-Ni(OH)$_2$ is slightly smaller than the ideal structure (8.0 Å), the space is still sufficiently wide enough for intercalation molecules like water (~0.27 nm) and nitrate (~0.326 nm) to reside with translational, vibrational, and rotational degrees of freedom without damaging the layers. Besides, to better probe the crystal phase of 2D thin flakes, an Ag-nanopillar-coated substrate (Fig. 3D inset), as a surface-enhanced Raman spectroscopy (SERS) substrate, is introduced for assistance since the synthesized 2D α-Ni(OH)$_2$ is too thin to be detectable by conventional Raman spectroscopy. A characteristic Raman peak of α phase is observed at 3610 cm$^{-1}$, as shown in Fig. 3D, which corresponds to the internal O-H stretching and is considered as a fingerprint for identifying α-Ni(OH)$_2$ from β-Ni(OH)$_2$ or the mixture.[14,61] Besides, the phonon mode of α-Ni(OH)$_2$ is observed at 460 cm$^{-1}$ and the 2$^{nd}$ order phonon modes of α-Ni(OH)$_2$ are observed at 979 and 1055 cm$^{-1}$, further proving the successful synthesis of α-Ni(OH)$_2$. Nonetheless, α-Ni(OH)$_2$ is undoubtedly the major phase in the bulk crystals, and further dominates in the thin flakes.

To test the thermal stability of the intercalated α-Ni(OH)$_2$, the bulk α-Ni(OH)$_2$ is heated up from room temperature to various temperatures up to 200 °C for 10 hours (Supplementary Fig. 7A).

PXRD spectra show no obvious change until 200 °C (Fig. 3C blue curve), suggesting the wide working window of synthesized α-Ni(OH)$_2$ (room temperature to 180 °C, purple curve). As for the solvent stability, optical images of a selected area covered by thin flakes, showing no difference after immersing in different solvent (isopropanol, acetone, ethanol and toluene) for 30 minutes (Supplementary Fig. 7B), suggesting the applicability of working in different solvent environments.

**Optical and Electronic Characterization of 2D α-Ni(OH)$_2$**

To investigate the optical and electronic properties of synthesized α-Ni(OH)$_2$, ultraviolet–visible spectroscopy (UV-Vis) is carried out on the Ni(OH)$_2$ thin flakes spin-coated on indium tin oxide (ITO) coated glass and the optical absorption spectrum (Fig. 4A) is obtained from 300 nm to 700 nm. In addition, Tauc plot is obtained (Fig. 4B) based on Tauc equation (Supplementary Eq. 1) to determine the optical bandgap. Two bandgaps, 2.54 eV and 3.51 eV, are extrapolated from two linear regions, which correspond to few-layers[62,63] and bulk Ni(OH)$_2$,[62,63] respectively. Note that the bandgap of 2D α-Ni(OH)$_2$ is smaller than that of its bulk counterpart, which is an exceptional deviation from the quantum size effect[62] and is opposite from most other 2D materials[64,65]. The absorption spectrum and Tauc plot of bulk α-Ni(OH)$_2$ in H$_2$O dispersion are shown in Fig. 4C and 4D, from which the bandgap of bulk α-Ni(OH)$_2$ is extracted as 3.51 eV, agreeing with the result measured on ITO glass and previous experimental literature[51]. Moreover, the band gap of the obtained Ni(OH)$_2$ is also measured using reflected electrons energy loss spectroscopy (REELS), which is a widely used technique that collects and measures the reflected electrons with different kinetic energy loss due to induced electronic transitions. As shown in Fig. 4E, suggesting a band gap of 2.52 eV by measuring distance between the center of elastic scattering peak and the cutoff of the low energy loss region, which agrees well with the 2.54 eV bandgap obtained from the absorption spectrum.

Combining with its large domain size, the relatively large bandgap of 2D α-Ni(OH)$_2$ motivates us to investigate its dielectric properties. An Au-Ni(OH)$_2$-Ti sandwich-like device is fabricated. Optical images of a single 2D Ni(OH)$_2$ flake on an Au electrode and the final device are shown in Supplementary Fig. 8A-B, respectively. Supplementary Fig. 7C shows the typical I-V curves of the devices of different thicknesses from 4.9 to 18.3 nm, with top Ti electrodes grounded and voltage applied on the bottom Au electrodes (Supplementary Fig. 8C inset). For all four devices, the current grows exponentially with positive voltage, while a much lower or negligible current is observed with negative voltage. This unipolar conducting characteristic is attributed to the difference in Schottky barrier height (S.B.H.) at the Au-Ni(OH)$_2$ and Ti-Ni(OH)$_2$ interfaces. Given a much higher work function of Au compared to Ti, Au is expected to form a higher barrier with 2D Ni(OH)$_2$. When a positive voltage is applied on the Au electrode, this barrier is forward biased, resulting in an exponential I-V curve once the voltage exceeds its threshold value. Otherwise, the barrier is reversely biased, and no current is present before breakdown. We further convert the bias voltage into electric field strength to extract the breakdown field of each device (Supplementary Fig. 9A-B), which is 4.47, 2.92, 1.73 and 1.64 MV/cm for 4.9, 8.6, 14.8, and 18.3 nm Ni(OH)$_2$ flakes, respectively (Supplementary Fig. 9C). The reduced dielectric strength at higher thickness matches the trend of other dielectric materials (e.g., h-BN),[66] which is attributed to the formation of multiple grains and defects during the synthesis process for thicker samples. Nonetheless, all the four tested flakes meet the requirement for gate dielectric layers of field effect transistors (FETs), as the leaking current is below 10 A/cm$^2$ at a field strength of 2.5 MV/cm.[67] The results demonstrate the great continuity and uniformity of the synthesized 2D Ni(OH)$_2$ flakes and indicate that they are promising candidates as dielectric materials in micro-nanoelectronics devices.

**Mechanism Investigation of 2D α-Ni(OH)$_2$ Synthesis**

To understand and rationalize the realization of 2D micrometer-sized domain crystals of Ni(OH)$_2$, we propose a mechanism based on edge-on condensation of [M(H$_2$O)$_6$]$^{2+}$ into the existing crystal nucleation center in solution, supported by the comprehensive study of the key factors in the synthesis to be discussed in the following section. Our mechanistic discussion consists of two parts, as presented in Fig. 5A: (1) edge condensation at the atomic level and (2) thermodynamics of macroscopic nucleation and growth rates.

On the atomic level, charge neutrality is an important driving force for the stability of single crystalline nickel hydroxide, where in-plane bonds have strong ionic character. Seed crystal formed by a few Ni$^{2+}$ ions with edge hydroxyl ions (Fig. 5A I) are essential for 2D intra-planar growth.[68] In order for a single crystal to maintain electrical neutrality, Ni$^{2+}$ ions forming the crystal edge will be coordinated by "hanging" hydroxyl groups and the buildup of localized negative charge incurred by edge hydroxyl ions will attracts more Ni$^{2+}$ ions to approach (Fig. 5A II). Meanwhile, the "hanging" water of solvated [Ni(H$_2$O)$_6$]$^{2+}$ units formed in aqueous solution will be substituted by hydroxyl groups rapidly in the solution since the octahedral Ni$^{2+}$ is a stable yet labile species. Once this association has been established, continued substitution will occur at the newly bound metal ion (Fig. 5A III) until it is pulled down into the crystal structure (Fig. 5A IV). The terminal H$_2$O ligands further deprotonate to form new hanging hydroxy groups and bind with new approaching [Ni(H$_2$O)$_6$]$^{2+}$ units (Fig. 5A V). Since the surface OH$^-$ of seed crystal are already bonded with three Ni$^{2+}$ and saturated, [Ni(H$_2$O)$_6$]$^{2+}$ units tend to approach to these intra-planar "hanging" ligands at the edge that have strong localized negative charge than perpendicularly to the plane with OH$^-$ ions that have incorporated into the crystal structure, the growth expands along the in-plane direction, leading to the formation of 2D structures.

In order to maximize 2D domains, we thus aim to amplify the rate of crystal growth over nucleation. Higher nucleation rates lead to smaller, finer particle sizes and more quickly consumes the precursors, such that $Ni^{2+}$ ions are not sufficient to coordinate to existing nuclei and grow to larger crystals. With slower nucleation, the $Ni^{2+}$ ions stay dissolved as $[Ni(H_2O)_6]^{2+}$ units, such that they can arrange into the more charge neutral 2D configuration, without prematurely crashing out as polycrystalline or amorphous solid with bulk morphology. In an idealized case where nucleation is minimized, promoting growth in the 2D plane is the next target. Out-of-plane growth are usually dictated by premature stacking of layers, or more random growth orientation leading to amorphous precipitate, where crystalline domains are small and arbitrarily directed. The thermodynamic conditions provided by high temperature and long reaction time allow the opportunity to aggregate ions to reach their more charge neutral 2D configuration.

**Investigation of Key Parameters in 2D α-Ni(OH)₂ Synthesis**

With the above understanding, we aim to engineer a hydrothermal recipe to produce large domains of ultrathin 2D α-Ni(OH)₂ which have never been achieved. Our strategy emphasizes on promoting *ab* in-plane isotropy, which involves systematic control of parameters including cooling rate, soaking temperature, starting pH, as detailed in the following.

We first investigate the cooling rate on the morphology of the Ni(OH)₂. We performed the synthesis at 120 °C under the cooling rate of 0.5, 1.5 and 3.0 °C/min. At 3.0 °C/min, we observed many random shaped flakes with rough surfaces (Supplementary Fig. 10A), on which islands and aggregates formed. As the cooling rate decreases to 1.5 °C/min, the best results are observed (Fig. 5B i), where circular thin flakes are easily found with average domain size of ~20 μm, an exciting new record for 2D LTMHs synthesis. Interestingly, when the cooling is further slowed down to 0.5 °C/min (Fig. 5B ii), the average domain size of obtained flakes is significantly smaller (around

1-2 µm). The reason behind is that when cooling rate is decreased, the precipitation of seed crystals is largely slowed down and the crystal growth is significantly encouraged, edge-on condensation happens rapidly on very limited amounts of seed crystals, overcoming the surface saturation of $OH^-$ and leading to a disorder add-on, eventually built 3D crystals and limited the size on intraplanar dimension. Moreover, PXRD results show no distinguishable difference among the $Ni(OH)_2$ synthesized from different cooling rate (Fig. 5B iii), suggesting cooling rate only subtly changed domain size and surface morphology without altered the crystal structure.

We further investigate the effect of the pH since $Ni(OH)_2$ is reported preferably formed and stable in the range of 9-13.[69] The sealed autoclave vessel used in the hydrothermal approach prevents the *in situ* monitoring of pH over the reaction, we only control the initial pH of the solution and evaluate trends in outcomes. The initial pH value is adjusted from 5.57 to 7.60 by adding potassium hydroxide (KOH) while the cooling rates is fixed as 1.5 °C/min and the reaction temperature as 120 °C (Supplementary Fig. 11). The optical images show that the optimal 2D morphology is produced when no KOH is added, which reliably corresponds to an initial pH of 5.8 (± 0.2). With raised pH, thin circular flakes decrease in size and abundance (Fig. 5C), until by pH=7.60, only small $Ni(OH)_2$ particles arranged into amorphous films are observed (Supplementary Fig. 11C). Without extra KOH addition, decomposition of urea provides gentle conditions in which $OH^-$ anions are generated steadily, promoting crystal growth and reducing the chance of nucleation. In this way, the solvated $[Ni(H_2O)_6]^{2+}$ have more time to spend in solution and have more opportunity to aggregate to existing nuclei and crystals (Fig. 5A II), rather than rapidly precipitating as particles. In contrast, the extra addition of $OH^-$ from KOH reduced the time to reach the "critical pH", at which point the solution is supersaturated with $Ni^{2+}$ and $OH^-$ species, thus significantly promoted fast precipitation of solid $Ni(OH)_2$ instead of intra-planar growth, which is not favorable to form

2D crystals. Furthermore, the introduction of extra OH⁻ species slows down the urea hydrolysis, resulting in less production of ammonia, which is less promotable to form α phase since ammonia has widely been identified as an intercalation species in $Ni(OH)_2$[51]. The PXRD clearly show a transformation from α-phase to polycrystalline to β-phase $Ni(OH)_2$ when increasing the pH (Fig. 5C ii). With higher starting pH, quenched d (001) at 2θ=12.5 ° from α-phase (grey arrow) and increased d (001) at 2θ=15.1 ° from β-phase (purple arrow) are observed, along with the disappearance of α-(301) and emergence of β-(100), β-(101) (green arrows) at 2θ=33.4 °, 38.2 °.[70] The increased integral intensity of (001) plane from β-phase implies the in-plane growth, however, smooth 2D morphology is not observed upon transition to the β-phase.

Besides, we also modulate the pH by ammonia hydroxide ($NH_4OH$), where low reaction yield is observed with extra $NH_4OH$. This is because the increased solvation of $Ni^{2+}$ ions by aqueous ammonia displaced water to form $[Ni(NH_3)_6]^{2+}$ (Fig. 5C iii) or monohydrate substituted $[Ni(H_2O)_x(NH_3)_{6-x}]^{2+}$ (where x≥1) instead of $[Ni(H_2O)_6]^{2+}$. Since the deprotonation is much harder for $NH_3$, compared to that of $H_2O$, the new seed crystals loses attraction to $[Ni(H_2O)_6]^{2+}$ units (Supplementary Fig. 12), which prevents 2D expansion based on our proposed mechanism described above. This hypothesis is further supported by the UV-Vis measurements. As shown in Supplementary Fig. 12G, the $^3A_{2g} \rightarrow ^3T_{2g}$ transition peak is found shifting to a higher energy, suggesting the presence of $[Ni(NH_3)]^{2+}$ due to larger energy splitting[71]. The solutions are also increased blue in intensity, characteristic to the hexamine $Ni^{2+}$ species. More information on control of $NH_4OH$ can be found in the Supplementary Fig. 13.

Another important factor we believe is the soaking temperature. We perform the synthesis at various temperatures from 80 °C to 100 °C, 120 °C, 140 °C and 160 °C (Fig. 5D and Supplementary Fig. 14), where the cooling rate is set as a constant, 1.5 °C/min. At 80 °C, only amorphous and bulky aggregates are observed (Supplementary Fig. 14A), which persists when the temperature is elevated to 100 °C (Supplementary Fig. 14B). We believe that under low temperature, the autogenous pressure generated in the fixed volume is too low to provide high supersaturation for $[Ni(H_2O)_6]^{2+}$ units to form and add onto the seed crystals. Accordingly, as the temperature increases to 120 °C, large circular flakes are observed (Fig. 5B i). However, at even higher temperature, 140 °C, surface defects, holes, and crimps are more widely observed in the optical image (Fig. 5D i), which is a consequence of promoted urea hydrolysis, resulting in increased production of free ammonia. The excess ammonia form $[Ni(NH_3)_6]^{2+}$ complexes with solvated $Ni^{2+}$, effectively consuming $Ni^{2+}$ ions and preventing the formation of $[Ni(H_2O)_6]^{2+}$ units. As discussed before, $[Ni(NH_3)_6]^{2+}$ hardly can deprotonate to form new negatively charged edge, thus lose attraction to more units to add on, which leads to holes and surface defects. Notably, while the temperature is further increased to 160 °C, the average size decreases to ~5 μm and some bulky aggregates are found gathering around the circular flakes (Fig. 5D ii), which is another consequence of nuclei edge blocking. The higher the temperature is, the more $[Ni(NH_3)_6]^{2+}$ and less $[Ni(H_2O)_6]^{2+}$ complexes are formed, then the more terminal groups on the existing nuclei and crystals are occupied by $[Ni(NH_3)_6]^{2+}$, thus edge substitution cannot proceed to achieve intra-planar growth, leading to smaller crystal size eventually. PXRD spectra of samples synthesized under different temperatures is shown in Fig. 5D iii. The peak from (100) at 18.3 ° (indicated by the green arrow) decreases significantly when increasing the temperature from 120 to 160 °C, suggesting intralayer growth along (100) direction is diminished. This result matches with the decreasing average domain size observed under optical images as increasing temperature. Moreover, as the temperature increased, the β (001) peak at 21.7 ° (grey arrow) becomes smaller and less visible, indicating the decreased proportion of β phase space in the α/β-interstratification structure, which is the consequence of interlayer swelling under high temperature, suggesting high temperature would produce purer and highly ordered α-Ni(OH)$_2$. Besides, we also observed a set of small peaks (purple arrow) along the 140 °C spectrum (Supplementary Fig. 14F) from 30 to 80 °, suggesting the disorder and rearrangement of the crystal structure due to an ongoing phase transition at this medium temperature.

In a nutshell, supported by the investigation of key parameters on the synthesis and the comprehensive characterizations of the resulting samples, the synthesis mechanism is constructed and proved involving the formation of seed crystals and [Ni(H$_2$O)$_6$]$^{2+}$ units, followed by continuous water and hydroxyl ligand-substitution, reaching supersaturation under moderate rate and pressure with suitable element ratio is the key to promote 2D intraplanar crystal growth.

**Conclusion**

In conclusion, we present a concept of directing the 2D growth of LTMHs through the spin-state controlled crystal field geometry, leading to the achievement of the largest ever 2D Ni(OH)$_2$ (>20 μm) and Co(OH)$_2$ (>40 μm) domains down to a few nanometers. A direct relationship between spin-state controlled geometry and macroscopic morphology is built, illustrating the 2D growth preferred geometry. The detailed characterization on the 2D Ni(OH)$_2$ shows the high crystallinity, α phase, and large bandgap of 2.5 eV, suggesting its great potential as dielectric materials in 2D electronic devices. In addition to the control of nucleation and growth rate, a 2D growth mechanism via edge-on condensation is established and tested by systematic investigation on the influence of key parameters. The reaction-morphology relationship elaborated herein provides a blueprint for a well-guided approach towards the successful design and development of more 2D LTMHs. Considering the great scalability of the hydrothermal synthesis method, our work suggests a promising future of joining LTMHs in the 2D family as a group of important members for future micro-nanoelectronics.

## Methods

### Hydrothermal Synthesis

The synthesis (schematic is shown in Supplementary Fig. 15) was initiated by combining 0.504 g nickel nitrate hexahydrate, an inexpensive, commercially available salt, with 0.456 g carbamide, commonly known as urea, in the mixture of ethylene glycol (28 ml) and deionized water (4 ml). After 30 minutes stirring under room temperature, the solution was transferred to a 100ml Teflon vessel, which was sealed into a hydrothermal autoclave reactor afterwards. For an optimal synthesis, the reactor was sent into an oven, in which the temperature was raised up to 120 °C at a speed of 1.5 °C/min. After 18 hours soaking, the reactor was cooled down to room temperature as the same speed as heating up, then the solution was washed by ethanol and deionized (DI) water for three times each and eventually dispersed in DI water for storage. The obtained flakes are high yielding on gram scales and gradually sinking to bottom over time.

### Sample Preparation

The synthesized thin $Ni(OH)_2$ flakes were extracted from the bulk material by centrifugation. After 4 hours centrifuge at 9500 rpm, the clear supernatant was drop-casted on: (1) 300 nm $SiO_2$/Si substrates for optical images, AFM, SEM, (2) Si substrates for XPS, (3) copper grids and silicon nitrite grids for TEM, (4) Ag-nanopillar-coated substrate for SERS, and dried by heating on hot plate at 80 °C for 2 hours. The clear supernatant and dispersion were spin-coated on ITO glasses for UV-Vis measurements. The sediment was collected and dried at 80 °C for 12 hours for PXRD.

### Chemicals and Materials

All reagents were purchased from Fisher chemical and used without further purification, with exception of nickel nitrate hexahydrate, which was recrystallized from supersaturated aqueous solution.

**Characterization Techniques**

Optical images were collected by optical microscopy (Nikon DS-Ri2). AFM topography was acquired on a Bruker Dimension 3000 in a tapping mode. SEM images were taken on Zeiss Supra 55. XPS spectra were collected from PHI Versaprobe II. TEM measurements were performed on a FEI Tecnai Osiris TEM, operating at a 200 keV accelerating voltage. SAED was measured on a JEOL 2100 TEM. The SAED simulation was performed through STEM_CELL software. PXRD scanning were performed on a Bruker D2 Phaser with Cu Kα radiation of wavelength $\lambda = 1.54184$ Å at 30 kV and 10 mA. SERS measurement was performed on a Horiba-JY T64000, using a triple-grating mode with 1800 g mm$^{-1}$ gratings, and a 532 nm laser line. Solution- and solid-state UV-Vis spectra were taken with Agilent Cary 5000. REELS measurement was carried on Thermo Scientific Nexsa G2 and data was analyzed by Avantage.

**Acknowledgement**

**Funding:** Acknowledgement is made to the donors of the American Chemical Society Petroleum Research Fund for support of this research. X.L. acknowledges the support of Boston University and Boston University Photonics Center. T.L. and J.C. acknowledge the support from the U.S. Department of Energy (DOE), Office of Science, Basic Energy Sciences (BES), under Award DE-SC0021064. L.P. and X.L. acknowledges the support of Ignition Award from Boston University and the ACS Petroleum Research Fund from American Chemical Society. H.Z.G. acknowledges the support of BUnano Cross-Disciplinary Fellowship. **Author contribution:** L. P. and X. L. conceived and designed the experiments. L. P., G. M., performed all the synthesis. L. P., H. G., T. L., J. C. and H. K. characterized the samples. L. P., G. M. and X.L. analyzed the data and wrote the manuscript. All authors discussed the results and commented on the manuscript. **Competing interests**: The authors declare that they have no competing interests. **Data and materials availability**: All data needed to evaluate the conclusions in the paper are present in the paper and/or in the Supplementary Information. Additional data related to this paper may be requested from the authors.

# Figures

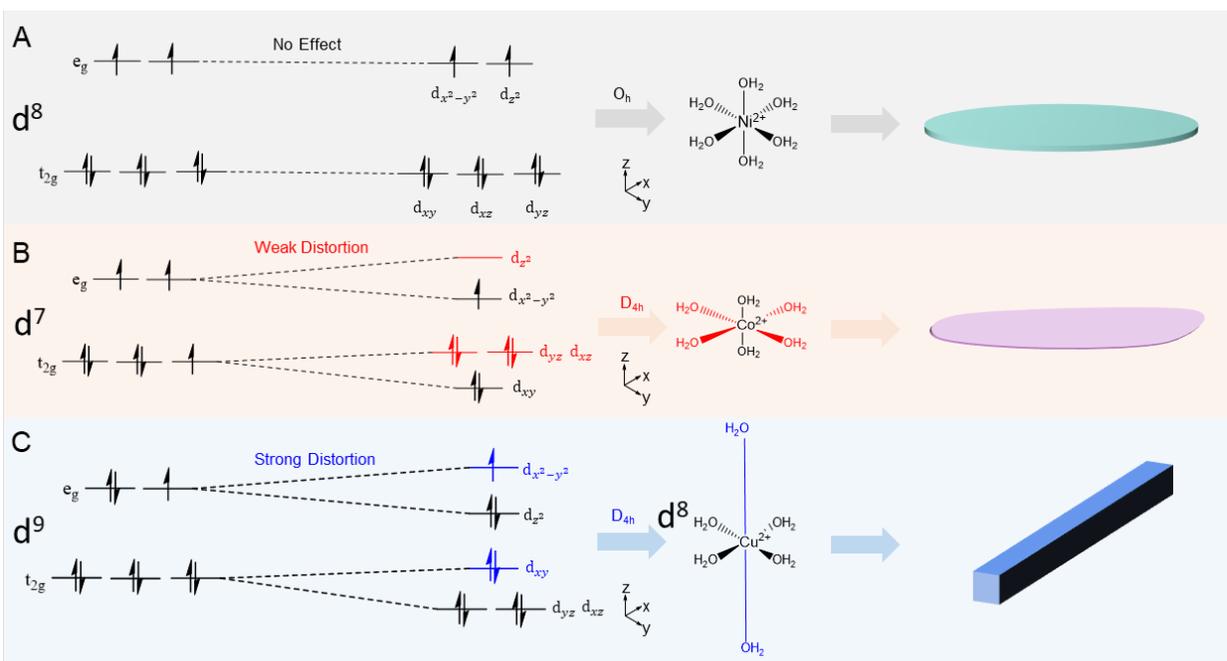

**Fig. 1.** *d*-orbital splitting diagram and spin-state directed geometries of six-coordinate $[M(H_2O)_6]^{2+}$ species with resultant morphology of $M(OH)_2$ crystals, [M=Ni, Co, Cu]. (**A**) $[Ni(H_2O)_6]^{2+}$ has perfect $O_h$ symmetry and yields the largest 2D crystalline flakes with circular morphology. (**B**) $[Co(H_2O)_6]^{2+}$ weakly distorts to $D_{4h}$ symmetry with weak compression along the z-axis, yielding fewer observed 2D flakes with irregular shape and varying thickness. (**C**) $[Cu(H_2O)_6]^{2+}$ distorts to $D_{4h}$ symmetry with dramatic elongation along the z-axis, yielding bulk crystals with prism-like morphology.

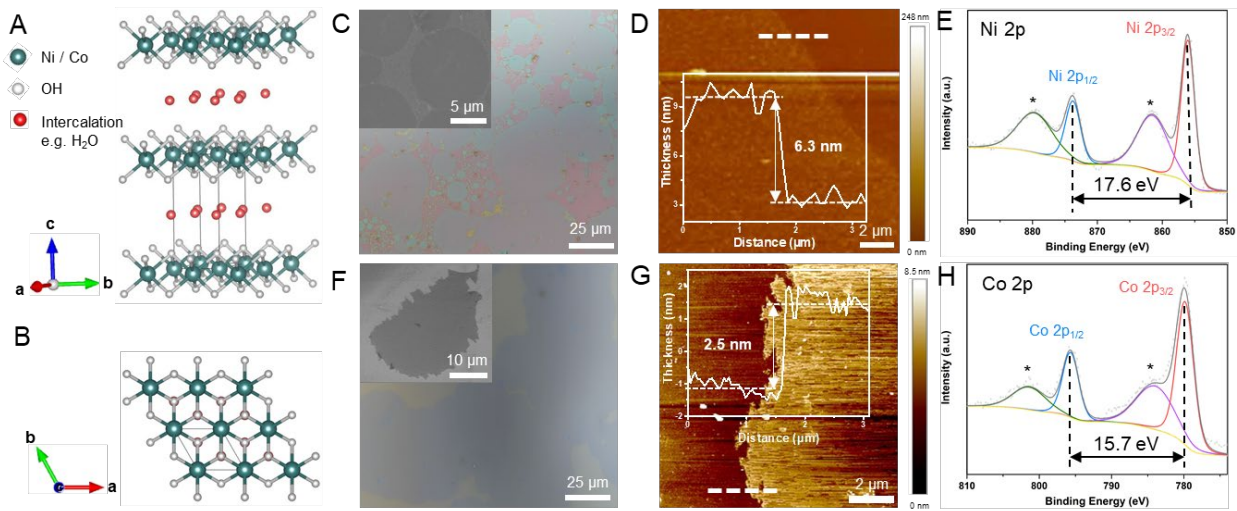

**Fig. 2.** Crystal structure and morphology characterization of Ni(OH)$_2$. **(A and B)** Crystal structure of α-M(OH)$_2$ (M=Ni and Co). (**A**) Side view; (**B**) Top view. (**C**) Optical image and SEM image (inset) of synthesized Ni(OH)$_2$ flakes on a SiO$_2$/Si substrate. (**D**) AFM image of a Ni(OH)$_2$ flake. The white dash line indicates the location where the thickness is measured. Inset: height profile along the white dash line. (**E**) XPS spectra of Ni 2p, suggesting a 17.6 eV spin energy separation, which is characteristic of the Ni$^{2+}$ in Ni(OH)$_2$. (**F**) Optical image and SEM image (inset) of synthesized Co(OH)$_2$ flakes on a SiO$_2$/Si substrate. (**G**) AFM image of a Co(OH)$_2$ flake, the white dash line indicates the location where the thickness was measured. Inset: height profile along the white dash line. (**H**) XPS spectra of Co 2p, suggesting a spin energy separation of 15.7 eV. Peaks labeled by "*" in (**E**) and (**H**) represent the satellite peaks generated by the secondary electrons during XPS measurements.

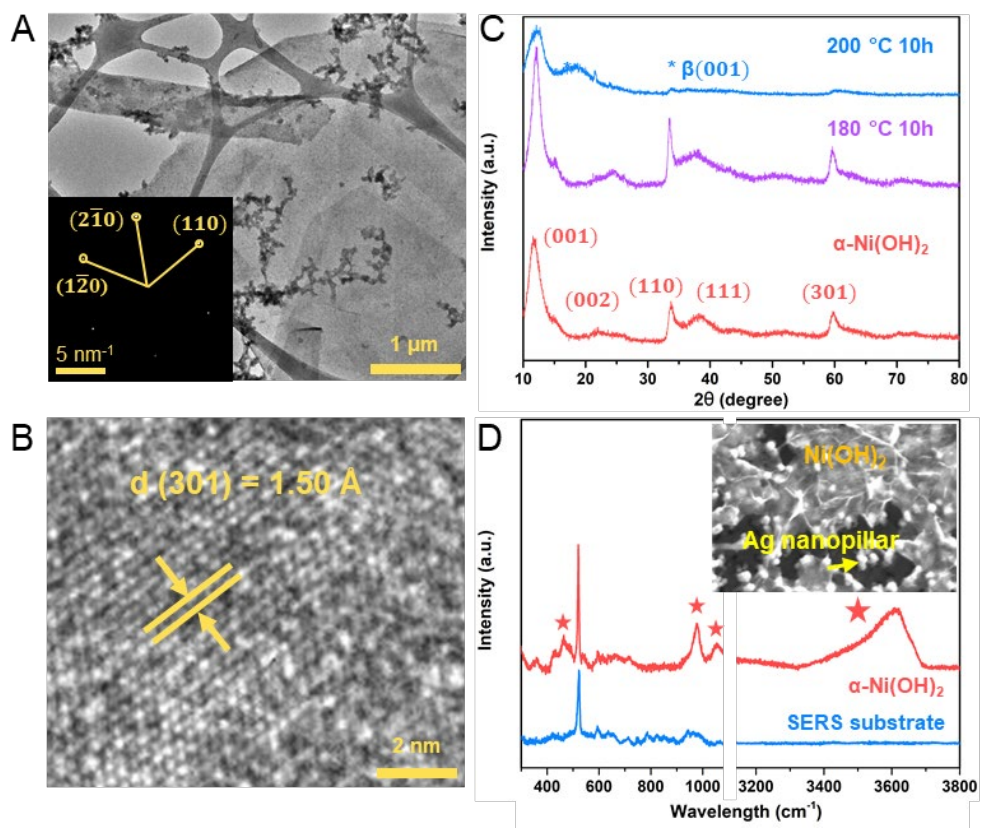

**Fig. 3**. Crystallographic characterization of Ni(OH)$_2$. (**A**) Low-magnification TEM image of Ni(OH)$_2$. Inset: the SAED pattern taken by a 25 cm camera. (**B**) High-magnification TEM image of Ni(OH)$_2$, the d-spacing of (301) lattice plane was measured as 1.50 Å. (**C**) PXRD pattern of bulk Ni(OH)$_2$ (red curve), in which the 2D large domain was observed; yellow and blue curves are the PXRD spectra of bulk Ni(OH)$_2$ after annealing at 180 (purple) and 200 °C (blue) under argon for 10 hours, respectively. More spectra from different temperatures are shown in Supplementary Fig. 7. (**D**) Raman spectra of Ag nanopillar SERS substrate with and without Ni(OH)$_2$ thin flakes on. Characteristic peaks of α-Ni(OH)$_2$ indicated by "stars" are observed and a fingerprint one is observed at 3610 cm$^{-1}$. Inset: SEM image of Ag nanopillar (white) SERS substrate coated with Ni(OH)$_2$ thin flakes.

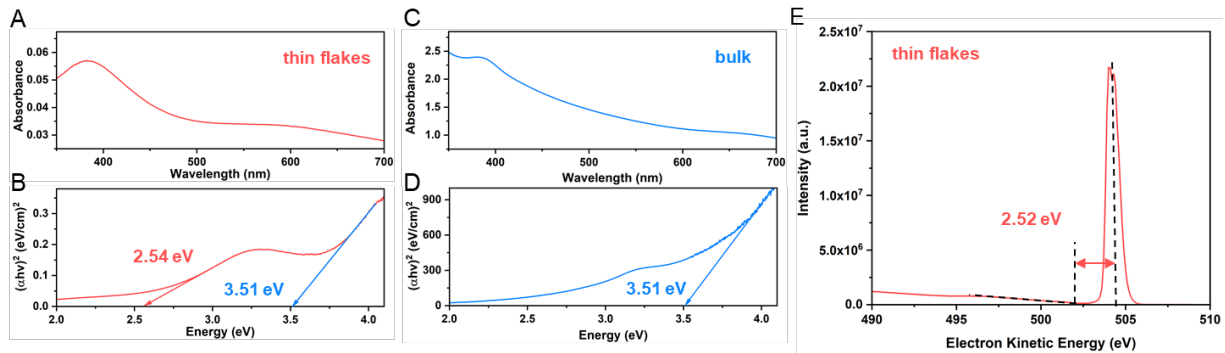

**Fig. 4**. Band gap measurement of Ni(OH)$_2$. (**A**) Optical absorption spectrum of few layers α-Ni(OH)$_2$ thin flakes on ITO glass, measured by UV-Vis from 300 nm to 700 nm. (**B**) Tauc plot derived from (**A**), the extrapolated bandgaps are 2.54 eV and 3.51 eV, corresponding to few layers and bulk α-Ni(OH)$_2$. (**C**) Optical absorption spectrum of bulk α-Ni(OH)$_2$ in H$_2$O dispersion, measured by UV-Vis from 300 nm to 700 nm. (**D**) Tauc plot derived from (**C**), the extrapolated bandgaps is 3.51 eV, corresponding to bulk α-Ni(OH)$_2$. (**E**) Band gap extrapolated from reflected electrons energy loss spectroscopy (REELS) by measuring the gap between the center of elastic scattering peak and the cutoff of the low energy loss region is 2.52 eV.

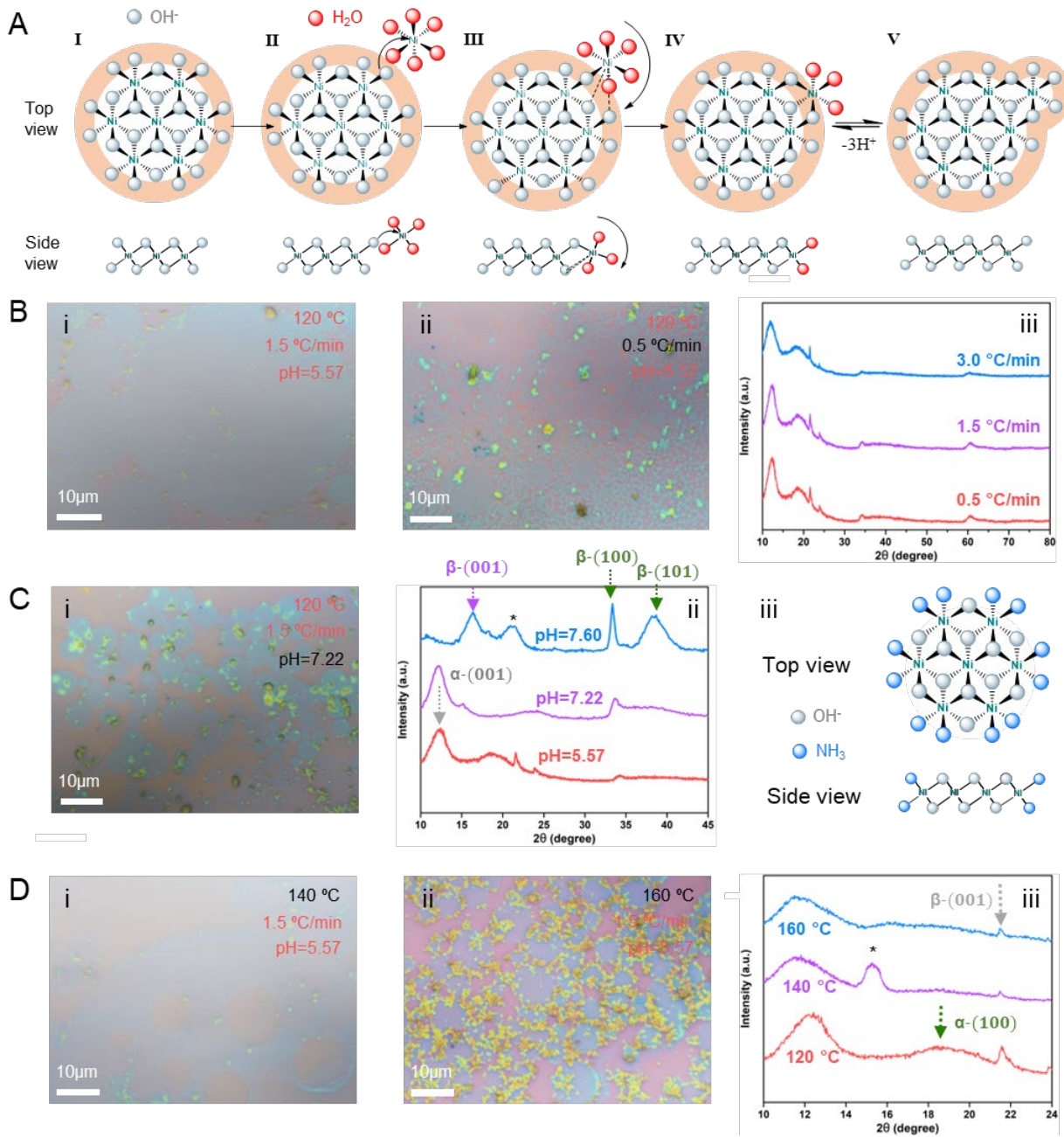

**Fig. 5**. 2D growth mechanism and parameters tuned synthesis of Ni(OH)$_2$. (**A**) 2D growth mechanism of Ni(OH)$_2$ via edge-on condensation after nucleation. **I**. The buildup of localized negative charge (orange ring) incurred by edge OH$^-$ ions on the nucleus attracts approach of Ni$^{2+}$ ions; **II**. Ligand exchange at labile Ni$^{2+}$ centers, undergoing substitution with hanging OH$^-$ groups when proximal to an existing crystal, which guides continues substitution of terminal OH$^-$ groups at newly bounded Ni$^{2+}$; **III**. The unit is locking into the crystal structure; **IV.** Deprotonation of coordinated H$_2$O or its displacement by OH$^-$ to extend the crystal domain (**V**) or to continue the add-on (**II**). (**B**) Optical images of α-Ni(OH)$_2$ synthesized under different cooling rate, i. 1.5 °C/min, ii. 0.5 °C/min; iii, PXRD spectra shows no crystal phase change. (**C**) Optical image of α-Ni(OH)$_2$ synthesized under different starting pH modulated by extra KOH. Increasing pH leads

to surface defects and smaller domain size (i) due to formation of $[Ni(NH_3)_6]^{2+}$, which terminates the 2D crystal expansion since it cannot deprotonate to form new negatively charged edge; (iii) PXRD spectra shows more β phase is formed with extra KOH addition. Peak labeled by "*" is generated from the stacking faults. (**D**) Optical images of α-Ni(OH)$_2$ synthesized under different soaking temperature, i. 140 °C with surface defects, ii, 160 °C with smaller domain size; iii, PXRD spectra show higher purity of α phase synthesized under higher temperature. Peak labeled by "*" suggests the sample is undergoing phase transition at this intermediate temperature. More details of parameters tuning can be found in Supplementary Fig 10-14.